\documentclass[epj,referee]{svjour}
\usepackage{graphics}
\usepackage{amsfonts}
\usepackage{epsfig}
\usepackage{epsfig,amsmath,amssymb,graphics,color,calc,cite,psfrag}
\begin{document}
\title{Spiral model, jamming percolation and glass-jamming transitions}
%\subtitle{un sacco!}
\author{Giulio Biroli \inst{1} \and Cristina Toninelli\inst{2}% etc
% \thanks is optional - remove next line if not needed
%\thanks{\emph{Present address:} Insert the address here if needed}%
}                     % Do not remove
%
%\offprints{}          % Insert a name or remove this line
%
\institute{ Service de Physique Th{\'e}orique, CEA/Saclay-Orme des Merisiers, F-91191 Gif-sur-Yvette Cedex, FRANCE \and Laboratoire de Probabilit{\'e}s et Mod{\`e}les Al{\'e}atoires CNRS UMR 7599 Univ. Paris VI-VII 4,Pl.Jussieu F-75252 Paris Cedex 05, FRANCE; e-mail ctoninel@ccr.jussieu.fr }
\date{Received: date / Revised version: date}
% The correct dates will be entered by Springer
%
\abstract{
The Spiral Model (SM) corresponds to a new class of kinetically constrained 
models introduced in joint works with D.S. Fisher \cite{letterTBF,TBFcomment}. 
They provide the first example of finite dimensional models with
an {\sl ideal glass-jamming transition}.
This is due to an underlying {\it jamming percolation} transition which 
has unconventional features:
it is discontinuous (i.e. the percolating cluster is 
compact at the transition) and the typical size of the clusters diverges faster than any power law, leading to a Vogel-Fulcher-like divergence of the relaxation
time. Here
 we present a detailed physical analysis of SM, see \cite{BTjsp} for rigorous proofs.\\
We also show 
 that our arguments for 
SM does not need any modification contrary to recent claims of Jeng and Schwarz \cite{JS}.
\PACS{64.70.Pf \and 05.20.-y \and 05.50.+q \and 61.43.Fs
%      {PACS-key}{discribing text of that key}   \and
%      {PACS-key}{discribing text of that key}
     } % end of PACS codes
} %end of abstract
\maketitle 
\section{Introduction}
Theoretical progress in understanding the glass and jamming transition, and more generally
glassy dynamics, is hampered by the shortage of finite dimensional models that display the basic phenomenological
ingredients and that are simple enough to be fully analyzed. Kinetically
Constrained Models (KCM) \cite{RitortSollich} are an exception.  
They have been introduced few decades ago \cite{FA,KA} as models for glass-forming liquids. 
They are based on the assumption that a particle does not (or cannot) move if surrounded by
too many others. This can be also interpreted in terms of dynamic facilitation \cite{GCPNAS}. 
%A lot of lattice models have been introduced in the literature. They all 
All KCM share two basics
properties: particles (or spins) can move (or flip) only if a certain constraint on the number
of occupied neighbors is verified. Once the constraint is verified the dynamical rules are such that
the resulting Boltzmann distribution is trivial, i.e. uncorrelated from site to site. As a consequence
the glass transition, if any, is purely dynamical in these models. Furthermore, another advantage is that 
the study of the dynamical transition can be reduced to a highly correlated percolation problem (of a new kind). In fact, in these models
a particle (or a spin) can be blocked if it has too many occupied (up) neighbors which can be blocked 
by their neighbors and so on and so forth. Using the fact that the Boltzmann equilibrium distribution
is trivial one can prove \cite{TB,TRoma} that the only dynamical transition that can take place
in these systems corresponds to a {\it jamming percolation} where an infinite cluster of mutually blocked
particles (spins) appears. In \cite{letterTBF,TBFcomment} we introduced
a new class of KCM which displays such a 
transition on a finite dimensional lattice, we  will thus refer to these models as Jamming Percolation (JP) models.
%Jamming Percolation (JP) models which display a
%glass-jamming transition on a finite dimensional lattice.
Here we will review the easiest example of a JP model, namely the
two-dimensional spin model which has been introduced in
\cite{BTjsp,TBFcomment} and dubbed \textsl{Spiral Model} (SM). For SM
the existence of a jamming transition 
has been rigorously proved \cite{BTjsp} 
and the exact value of $p_c$ has been identified: $p_c$ coincides with 
the critical threshold of directed site
percolation (DP) in two dimensions, $p_c^{DP}\simeq 0.705$. 
Contrary to recent claims \cite{JS} our proof for SM does not need any modification
(we will pinpoint in section \ref{Tjunctions} the incorrect assumption of  \cite{JS}). 
This jamming transition has remarkable properties: the density of the 
frozen cluster, $\Phi(p)$, is discontinuous at the transition but the cross-over length over which the 
system is still ergodic (or liquid) diverges. Furthermore the time scale for relaxation 
(and also the cross-over length) diverges faster than any power law. 
These properties are quite unusual but are exactly what is often assumed 
the real glass or jamming transition should display (if they exist).
They have been also rigorously proved in \cite{BTjsp}
%(see also \cite{TRoma}) 
modulo the standard conjecture
on the existence of two different correlation lengths for DP \cite{Hinrichsen}.

In the following we will sketch in a physical and hopefully transparent way 
the arguments which lead to the above results providing the tools needed 
 to analyze this transition and explaining the underlying mechanism:
it is the consequence of two perpendicular
directed percolation processes which together can form a compact
network of frozen directed paths  at criticality.
In the final section we shall discuss the generality of our 
approach and the universality of the jamming percolation transition of SM.

\section{The spiral model and its related percolation problem}

Consider a square lattice and,
for each site $x$, define among its first and second neighbours the couples of its
North-East (NE), South-West (SW), North-West (NW) and South-East(SE)
neighbours as in Fig.1,
%\ref{BPvoids}
namely NE$=(x+e_1,x+e_1+e_2)$,
SW$=(x-e_2,x-e_1-e_2)$, NW$=(x-e_1,x-e_1+e_2)$ and
SE$=(x+e_1,x+e_1-e_2)$, where $e_1$ and $e_2$ are the coordinate unit vectors. 
The spiral model is a stochastic spin lattice model where a spin can flip 
if and only if the following constraint is verified: 
both its NE \textsl{and/or} both its SW neighbours
are down \textsl{and} both
its SE \textsl{and/or} both its NW neighbours are down too (see Fig.1).
% \ref{BPvoids}
\psfrag{b}[][]{a)}
\psfrag{c}[][]{b)}
\begin{figure}[htp]
\includegraphics[width=0.99\columnwidth]{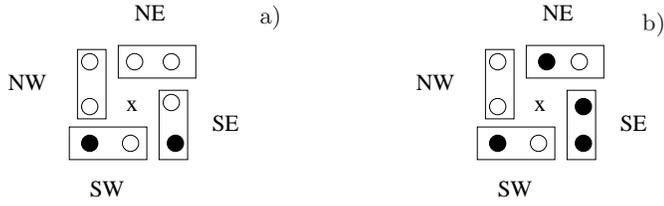}
\label{BPvoids}
\caption{Site $x$ and its NE, NW, SE and SW neighbours. Filled (empty) dots stand for up (down) spins. The constraint is (is not) verified at $x$ in case (a) (in case (b)).}
\end{figure}
If the constraint is verified then the spin flip rate is $p$ from down to up and $1-p$ from 
up to down. As a consequence, the invariant probability measure reached for $p<p_c$ is 
the Bernoulli product measure, i.e. independent from site to site. 
It is such that a spin is up with probability $p$ and down with probability $1-p$.
As explained in the introduction the dynamical transition takes place when 
an infinite cluster of mutually blocked spins appears with probability one with 
respect to the Bernoulli product measure. An easy way to unveil the existence 
of this cluster is to run the cellular automaton defined by the following 
update rule: down (up) spins are mapped to empty (occupied) sites; 
empty sites remain empty, occupied sites
get emptied only if the kinetic constrain is verified. 
The remaining cluster coincides with all spins up that are mutually blocked 
under the stochastic dynamics.  
Note that the kinetic rules can be also rephrased by saying that at least one
among the four sets $NE\cup SE$, $SE\cup SW$, $SW\cup NW$ and $NW\cup NE$
should be completely empty. From this perspective (and identifying an occupied site
with high density regions in a liquid) SM encodes in a very simplified way the 
cage effect that emerges in liquids and granular media close to the glass and jamming
transition.

\section{Critical density}

\subsection{Occurrence of blocked clusters for $p>p_c^{DP}$}
\label{Absence}

In order to establish $p_c<1$ we
will identify a set of blocked
clusters and show that they exist with probability one 
(with respect to the Bernoulli measure)
for $p>p_c^{DP}$, therefore $p_c\leq p_c^{DP}<1$.

Let us start by recalling the definition and a few basic results on DP
(see e.g. \cite{Hinrichsen}). Take a square lattice with randomly (independent) 
occupied sites and put two arrows going out
from each site $x$ towards its neighbours in the positive coordinate
directions, $x+e_1$ and $x +e_2$. 
On this directed lattice a continuous
percolation transition occurs at a non trivial critical density
$p_c^{DP}\simeq 0.705$ (a percolating cluster is now one
which spans the lattice following the direction of the arrows).
This transition is second order, as
for site percolation, but belongs to a different universality class. In particular, due
the anisotropy of the lattice, the typical sizes of the
incipient percolating cluster 
 in the parallel ($e_1+e_2$) 
 and transverse ($e_1-e_2$) 
directions diverge with different
exponents, $\xi_\parallel\simeq (p_c^{DP}-p)^{-\nu_{\parallel}}$ and
$\xi_{\parallel}\simeq \xi_{\perp}^z$ with $\nu_{\parallel}\simeq
1.74$ and $z\simeq 1.58$.  

Back to the Spiral Model, let us consider  the
directed lattice that is obtained from the square lattice putting two
arrows from each site towards its NE neighbours, as in
Fig.2 
%\ref{blocked} 
a). This lattice is equivalent to 
the one of DP,
simply tilted and squeezed. Therefore, for $p>p_c^{DP}$, there exists
a cluster of occupied sites which spans the lattice following the
direction of the arrows (cluster inside the continuous
line in Fig.2
% \ref{blocked} 
a)). We denote by {\sl{NE-SW clusters}} the occupied sets which follow the arrows of such lattice and {\sl{NW-SW clusters}} those that follow instead the arrows drawn starting from each site towards its NW neighbours. Consider now a site in the interior of a spanning
NE-SW cluster, e.g.site $x$ in the Fig.2
% \ref{blocked}
a): by
definition there is at least one occupied site in both its NE and SW
neighbouring couples, therefore $x$ is occupied and blocked with respect to the 
updating rule of SM. Thus, the presence of the DP cluster implies a blocked cluster
and $p_c\leq p_c^{DP}$ follows.  Note that these results would remain
true also for
 a different updating rule with
the milder requirement that only at
least one among the two couples of NE and SW sites is completely
empty (and no requirement on the NW-SE direction). 
However, as we shall see, the 
coexistence of the constraint in the NE-SW and NW-SE directions
is crucial to find a discontinuous transition for SM, 
otherwise we would have a standard DP-like continuous transition.

\psfrag{a}[][]{a)}
\psfrag{b}[][]{b)}
\psfrag{x}[][]{$x$}
\begin{center}
\begin{figure}[htp]
\includegraphics[width=0.9\columnwidth]{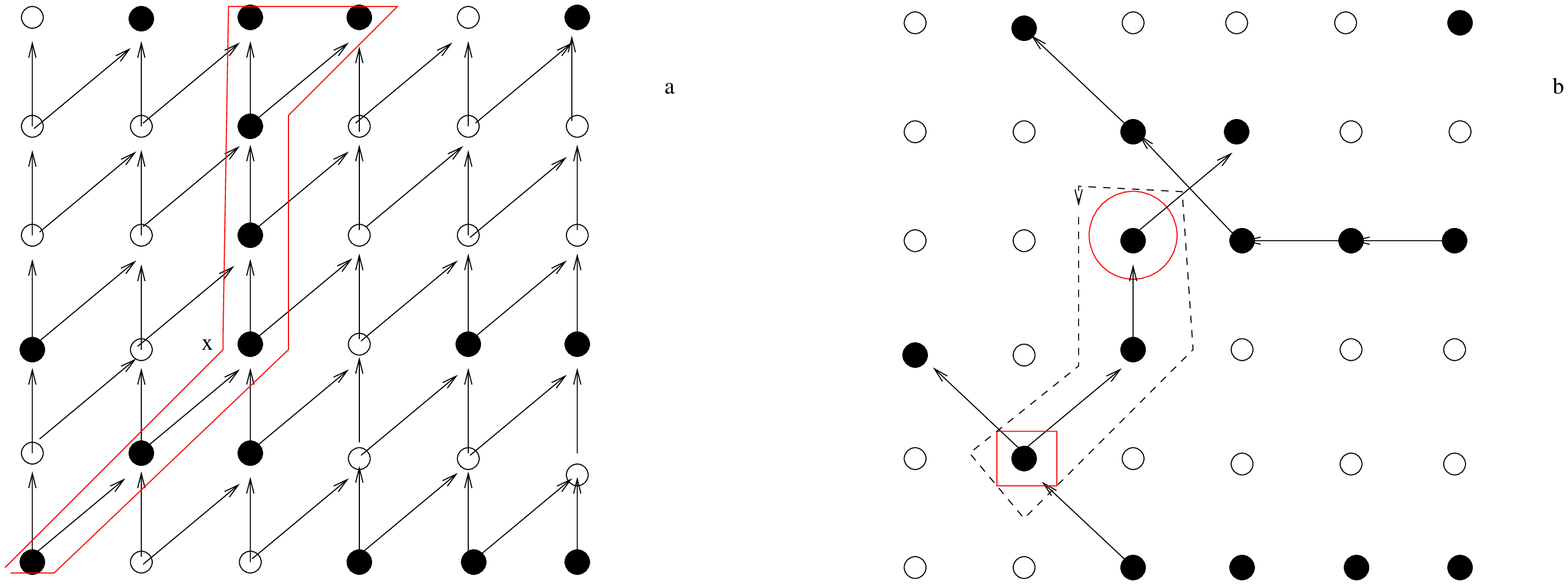}
\label{blockedfig}
\caption{a) The directed lattice obtained drawing arrows from each site towards its NE neighbours. Particles inside the continuous line belong to a NE-SW spanning cluster, thus they are blocked. b) A NE-SW non spanning cluster crossing two NW-SE clusters.  We depict both types of T-junctions: the bottom (top) one occurs with (without) a site in common for the NE-SW and NW-SE cluster. In both cases the last point of NE-SW before the crossing is blocked by the presence of the NW-SE cluster, thus all sites inside the dashed line cannot be erased before erasing at least one site of the NE-SW cluster.}
\end{figure}
\end{center}

\subsection{Absence of blocked clusters for $p<p_c^{DP}$}
%\label{Absence}

Before showing that below $p_c^{DP}$ blocked clusters do not occur,
 a few remarks are in order. If instead of SM 
 we were considering the milder rules described at the end of previous
 section, the result would follow immediately since the
 presence of a blocked cluster would imply the existence of a DP one.  On
 the other hand for SM rules, since blocking can occur along
 either the NE-SW or the NW-SE direction (or both), a directed path
 implies a blocked cluster  but the
 converse is not true.  This is because a NE-SW non spanning cluster
 can be blocked if both its ends are blocked by a T-junction with
 NW-SE paths, as  shown in Fig.2
% \ref{blockedfig} 
b) (see section \ref{Tjunctions} for 
 a detailed definition of T-junction). 
By using such T-junctions
 it is also possible to construct frozen clusters which do not contain
 a percolating DP cluster neither in the NE-SW nor in the
 NW-SE direction: all NE-SW (NW-SE) clusters
 are finite and are blocked at both ends by T-junctions with finite
 NW-SE (NE-SW) ones (see Fig.3
% \ref{blockedfig} 
b)).  As we will show in section
 \ref{discontinuity} these T-junctions are crucial to make the behavior
 of the transition for SM very different from DP, 
although they share
 the same critical density.  This also means that the fact that
 spanning DP clusters do not occur for $p<p_c^{DP}$ is not sufficient to
 conclude that also blocked clusters do not occur. What
 strategy could one use? Recalling bootstrap percolation results
\cite{bootstrap}, a possible idea is to
 search for proper unstable voids
 from which we can iteratively empty the
 whole lattice.
Of course, since we already know that blocked clusters occur
 when $p\geq p_c^{DP}$, something should prevent this unstable voids to
 expand at high density.

 \psfrag{s}[][]{{{$s$}}}
 \psfrag{a}[][]{{{\Huge{a)}}}}
\psfrag{b}[][]{{{\Huge{b)}}}}
\psfrag{s6}[][]{{{$s/6$}}}
 \psfrag{A}[][]{{{\tiny{A}}}}
 \psfrag{B}[][]{{{\tiny{B}}}}
 \psfrag{C}[][]{{{\tiny{C}}}}
 \psfrag{D}[][]{{{\tiny{D}}}}
 \psfrag{E}[][]{{{\tiny{E}}}}
 \psfrag{F}[][]{{{\tiny{F}}}}
 \psfrag{G}[][]{{{\tiny{G}}}}
 \psfrag{H}[][]{{{\tiny{H}}}}
\psfrag{L}[][]{\Large{$x_L$}}
\psfrag{ell}[][]{{\huge{$\ell$}}}
\psfrag{x}[][]{$x$}
%\psfrag{x}[][]{\huge{a)}}
\begin{center}
\begin{figure}[htp]
\includegraphics[angle=0,width=0.52\columnwidth]{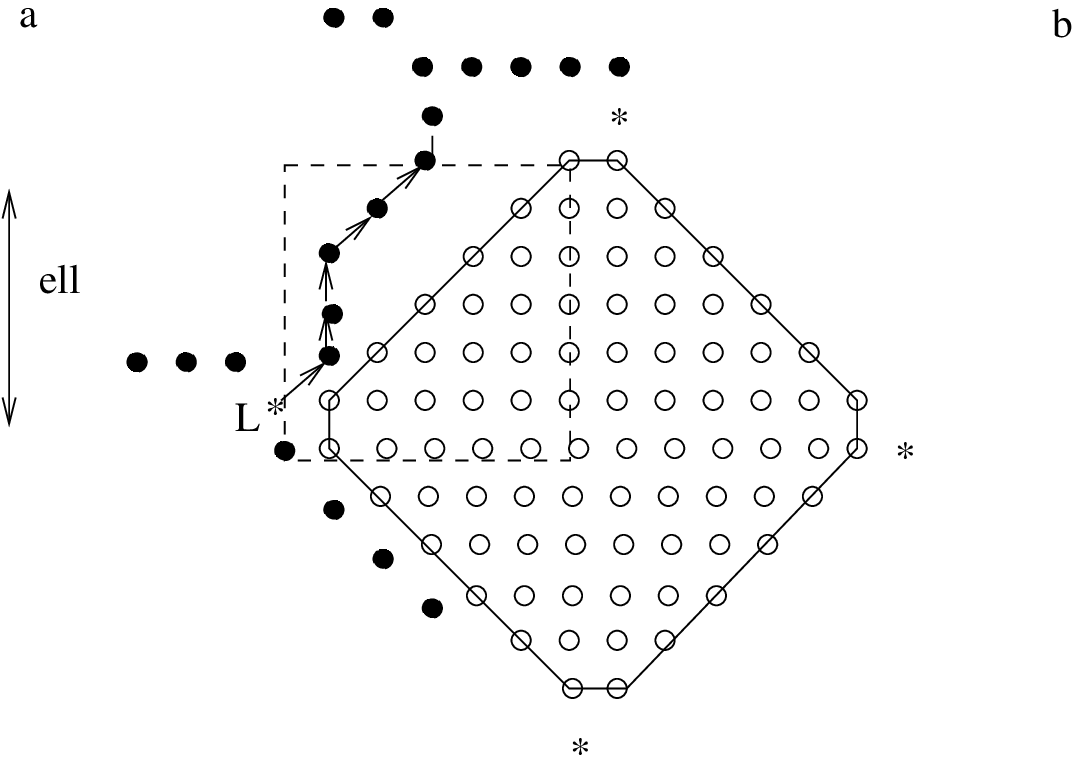}
\hspace{0.15 cm}
\includegraphics[width=0.43\columnwidth]{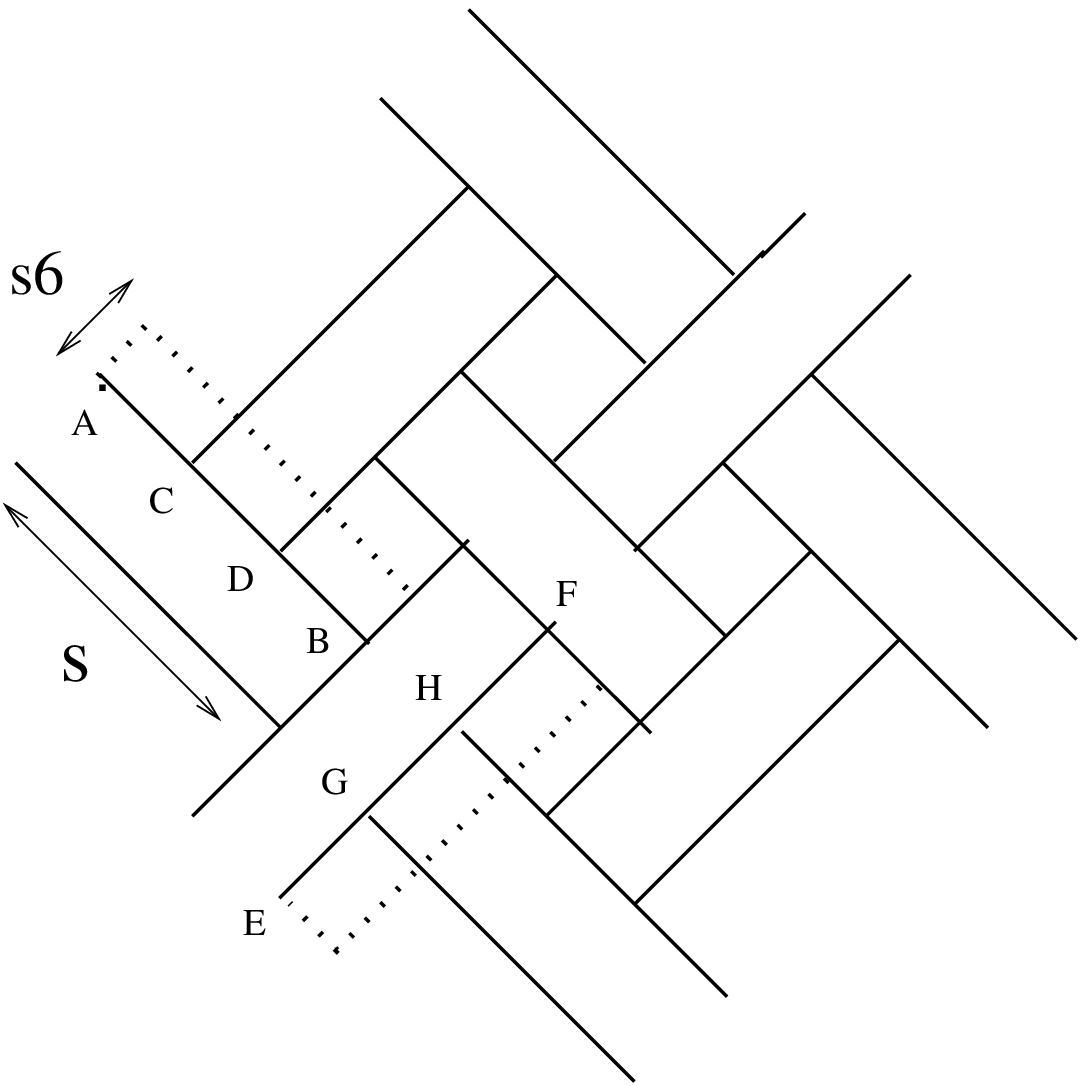}
\label{Qell}
\caption{a) $Q_{\ell}$ (continuous line),
$S_{\ell}$ (dashed line) and  the four sites ($*$) which should
be emptied to expand $Q_{\ell}$.
We draw the necessary condition for $x$ to be frozen: it should belong to a NE-SW cluster spanning $S_{\ell}$ and supported by NW-SE cluster from the exterior. b)The frozen structure described in the text: continuous lines
  stand for occupied NE-SW or SW-NE clusters.
 Each of these clusters is blocked since it ends in a
  T-junction with a cluster along the transverse direction. 
The dotted rectangle adjacent to cluster AB (EF) are the regions in which this
  cluster can be displaced and
  yet a frozen backbone is preserved: the T-junctions in C and D (G and
  H) will be displaced but not disrupted.}
\end{figure}
\end{center}

Consider the region $Q_{\ell}$
 inside the continuous line in Fig.3 a),
%\ref{Qell}
namely a ``square'' of linear
 side $\sqrt 2 \ell$ tilted of $45$ degrees with respect to the coordinate
 axis and with each of the four vertexes composed by two sites.  If
 $Q_{\ell}$ is empty and the four sites  external and
 adjacent to each vertex denoted by $*$  in Fig.3 a)
%\ref{Qell}
 are also empty, then it is possible to enlarge the empty region
 $Q_{\ell}$ to $Q_{\ell+1}$. Indeed, as can be
 directly checked, 
all the sites external to the top right side can be subsequently emptied
starting from the top  one and going downwards.
For the sites external to the other three sides of $Q_{\ell}$ we can
proceed analogously, some care is only required in deciding whether to
start from top sites and go downwards or bottom ones and go
upwards. Therefore we can expand $Q_{\ell}$ of one step provided all
the four $*$ sites are empty or can be emptied after some iterations
of the cellular automaton.  
Let us focus on one of these $*$ sites, e.g. the left one,
$x_L$ in Fig.3 a).
%\ref{Qell}. 
As it can be proved by an iterative procedure (see
\cite{BTjsp}), in order for $x_L$ not to be
emptyable there should exist a NE-SW cluster which spans the square $S_{\ell}$
of size $\ell$ containing the top left part of $Q_{\ell}$
(region inside the dashed line in Fig.3 a)).
%\ref{Qell})
This is due to the
fact that any directed path in the NW-SE
direction can be unblocked starting from the empty part of $S_{\ell}$ below 
the diagonal in the $e_1+e_2$ direction.
Therefore the only way in which $x_L$ can be blocked is that it belongs to a NE-SW cluster that  
is either supported by NW-SE clusters running outside $S_{\ell}$ or that is infinite. In any case
this  NE-SW cluster has to be at least of length $\ell$.
As a consequence, for large $\ell$, ($\ell >> \xi_{\parallel}$ ) the cost for a one step expansion of $Q_{\ell}$ is proportional to 
the probability of not finding such a DP path for any of the four * sites, 
$1-4\exp(-c\ell/\xi_{\parallel})$.   Thanks to the positive correlation among events at different $\ell$'s, the probability that the emptying
procedure can be continued up to infinity is bounded from below by the
product of these single step probabilities
%\footnote{The event
%that a directed spanning cluster does not occur when expanding from
%$\ell$ to $\ell+1$ and from $\ell+1$ and $\ell+2$ are not independent 
%since $S_{\ell}$ and $S_{\ell+1}$
% do intersect (and the same is
%true on the following steps). However these events are positively
%correlated, therefore the joint probability is bounded from below by
%the product of single step probabilities.} 
which goes to a strictly
positive value for $p<p_c^{DP}$ since $\xi_{\parallel}<\infty$. Note
that, as we already knew from the results of section 3.1,
this is not true for $p>p_c^{DP}$: the presence of long DP paths
prevents the expansion of voids. As a conclusion, the probability of emptying 
the whole lattice starting from an empty square $Q_{\ell}$ 
centered around a {\it given point} in the lattice and with  
$\ell >> \xi_{\parallel}$ is finite (although very small). 
Since there is an infinite number of points in the lattice, 
there will be at least one (actually a finite fraction) 
of sites from which the whole lattice can be emptied\footnote{Mathematically, one would say that the ergodic theorem
implies that in the thermodynamic limit with probability one the
final configuration is completely empty.} for 
$p<p_c^{DP}$. 
This, together with the result of section 3.1, 
yields $p_c=p_c^{DP}$.

\section{Critical behavior}
\subsection{T-junctions}
\label{Tjunctions}
One of the most important characteristic of the SM model, already 
alluded to in the previous sections, is that a directed path in the 
NE-SW direction can be supported by another path running in the 
NW-SE direction (and viceversa) via a T-junction. Let us discuss this point in detail
since recently it has been incorrectly claimed that this is not true for the SM
model. There are only two possible types of crossing of a NE-SW path with a 
NW-SE one:
%both 
%leading to what we call a T-junction: 
either they have one point in common or not. In the latter
case they should cross as in the upper crossing of Fig.2 b).   
%\ref{blockedfig} b).
In both cases we call the crossing a T-junction and the key observation is that if a NE-SW path ends in two T-junctions with NW-SE paths (or viceversa), it does not need to continue 
beyond the crossings in order to be blocked, as long as the NW-SE paths are 
blocked. If the T-junction occurs with a site in common (site inside the square of Fig.3 b)
%\ref{blockedfig}) 
this is a trivial consequence
of the fact that this point belongs to the NW-SE path. In the other case  it can be easily checked 
that the last point belonging to the NE-SW path (site inside the circle of Fig.2 b)
% \ref{blockedfig})
 is blocked thanks to 
the one above it, which belongs to the NW-SE path. All other possible 
crossings are related to these two cases by symmetry.  
In \cite{JS} it is stated that in order for a NE-SW path to stabilize 
a NW-SE path (or the converse) 
they shouldn't only cross but also have a point in common
and since this may not happen  
our proof for the SM needs a modification.   
As explained above this conclusion is incorrect and our proof for the SM model
does not need any modification (see also \cite{BTjsp} for further details). 

\subsection{SM: Discontinuity of the transition}
\label{discontinuity}

In the previous sections we have shown that
 the percolation
transition due to the occurrence of a frozen backbone for the Spiral
 Model occurs  at $p_c^{DP}$. We
will now explain why 
%it is
%qualitatively different from DP and any conventional
%percolation transition:
 the density of the frozen cluster  is
discontinuous, $\Phi(p_c^{DP})>0$ (the frozen structures are compact
rather than fractal at criticality). 
%and their typical size increases faster than any
%power law for $p\nearrow p_c^{DP}$.

By translation invariance $\Phi(p_c^{DP})$ is equal the probability 
that a given point, e.g. the origin, is blocked (i.e. 
it belongs to an infinite blocked structure). In order to show
that $\Phi(p_c^{DP})>0$ we will then  
construct a set of configurations, ${\cal{B}}$, for which the origin
is blocked and such that $P_{\cal{B}}(p_c^{DP})>0$. Since 
$\Phi(p_c^{DP})\ge P_{\cal{B}}$ our result implies $\Phi(p_c^{DP})>0$. 
In order to define $\cal{B}$, 
consider a configuration in which the origin belongs to a NE-SW path
of finite length $\ell_0/2$: this occurs with some finite probability $q_0>0$. 
Now focus on the infinite sequence of pairs of rectangles of increasing size
$\ell_i\times\ell_i/12$ with $\ell_1=\ell_0$, $\ell_i=2\ell_{i-2}$ and
intersecting as in Fig.4 a).
%\ref{disc} a)
A configuration belongs to
${\cal{B}}$ if {\sl each} of these
rectangles with long side along the NE-SW (NW-SE) diagonal contains a
NE-SW (NW-SE) percolating path (dotted lines in Fig.4 b)).
% \ref{disc} b)).
If this is the case then the infinite
backbone of particles containing the origin (cluster inside the 
continuous line in Fig.4 c))
% \ref{disc} c)) 
survives thanks to
the T-junctions among paths in intersecting rectangles. Therefore 
$\Phi(p)>q_o\prod_{i=1,\infty}P(\ell_i)^2$, where $P(\ell_i)$ is the
probability that a rectangle of size $\ell_i\times 1/12 \ell_i$ with
short side in the transverse direction is spanned by a DP cluster.
% \footnote{Again, since the rectangles do intersect the events that they are spanned are not independent. However these are positively correlated and the lower bound follows.}.
Recall that there is a parallel and a transverse length for DP with
different exponents, i.e. a cluster of parallel length $\ell$ has
typically transverse length $\ell^z$ \cite{Hinrichsen}. 
Let us divide the $\ell_i\times 1/12 \ell_i$
rectangle into  $\ell_i^{1-z}$ slices of size $\ell_i\times 1/12 \ell_i^{z}$.
For each slice the probability of having a DP cluster along the parallel
direction at $p_c^{DP}$ is order unity. Thus, the probability of {\sl
not} having a DP cluster in each of the slice is
$1-P(\ell_i)=O[\exp(-c\ell_i^{1-z})]$. From this  result
and the above inequality we get $\Phi(p_c^{DP})>0$.
Therefore the infinite cluster of jamming
percolation is ``compact'' with dimension $d=2$ at the transition.

\begin{figure}[htp]
\centerline{
\psfrag{ell412}[][]{$\ell_4/12$}
\psfrag{ell4}[][]{$\ell_4$}
\psfrag{l0}[][]{{{{$\ell_0$}}}}
\psfrag{r1}[][]{{{}}}
\psfrag{r2}[][]{}
\psfrag{r3}[][]{}
\psfrag{r4}[][]{}
\psfrag{r5}[][]{}
\psfrag{r6}[][]{}
\psfrag{r7}[][]{}
\psfrag{r8}[][]{}
\psfrag{a}[][]{a)}
\psfrag{O}[][]{\tiny{O}}
\psfrag{b}[][]{b)}
\includegraphics[width=0.27\columnwidth]{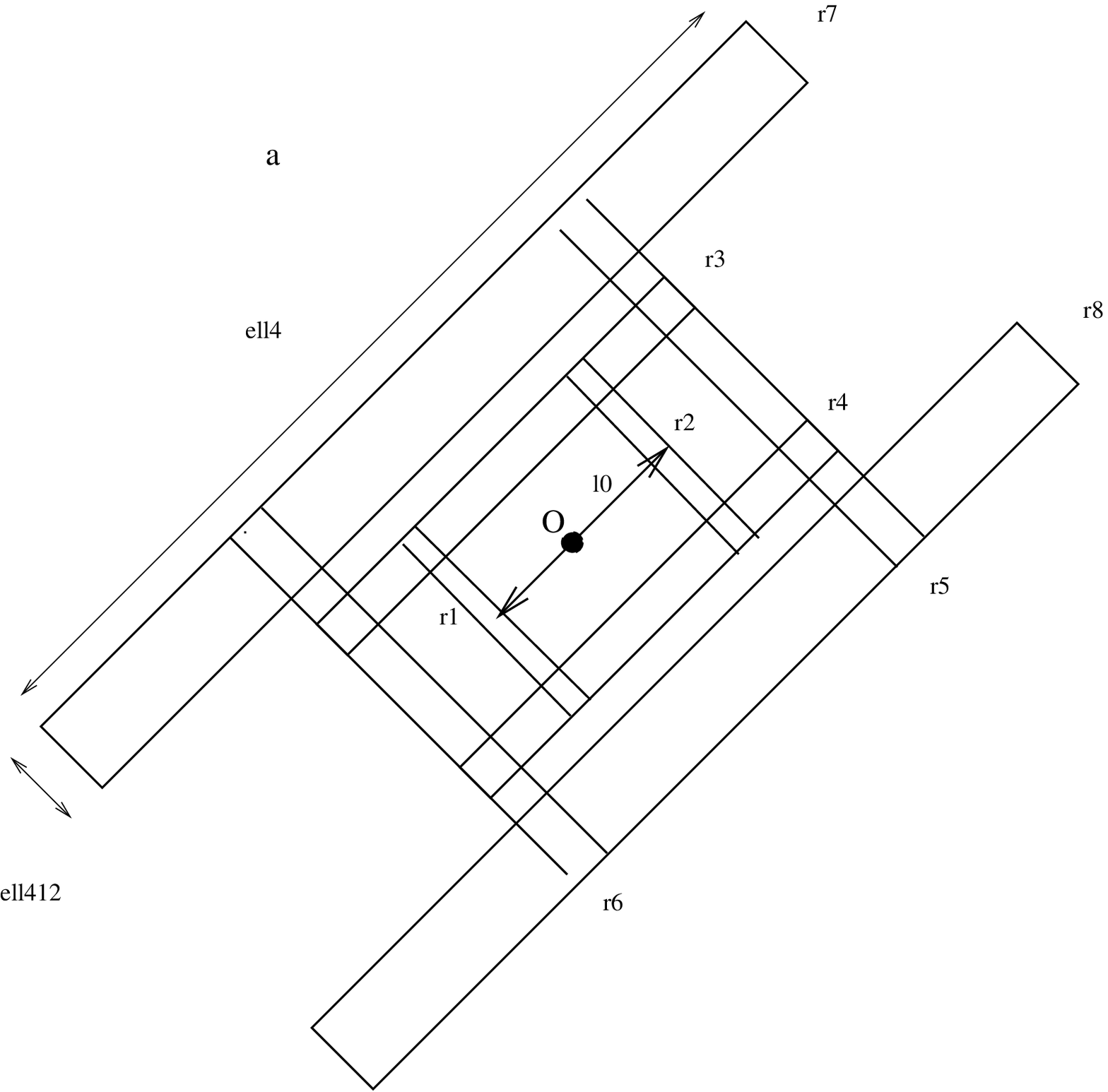}
\hspace{0.3 cm}
\includegraphics[width=0.27\columnwidth]{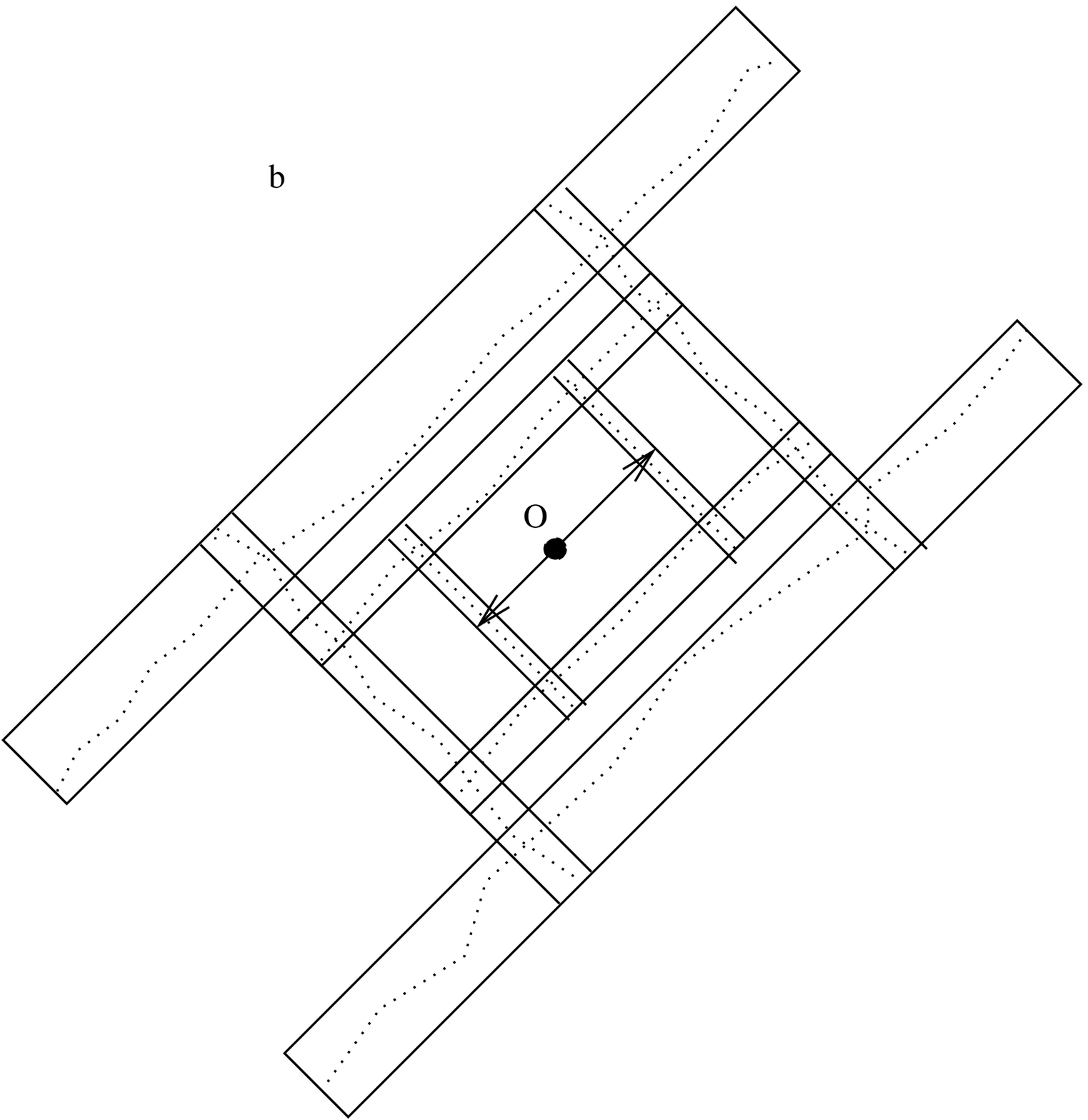}
\hspace{0.3 cm}
\psfrag{c}[][]{c)}
\includegraphics[width=0.27\columnwidth]{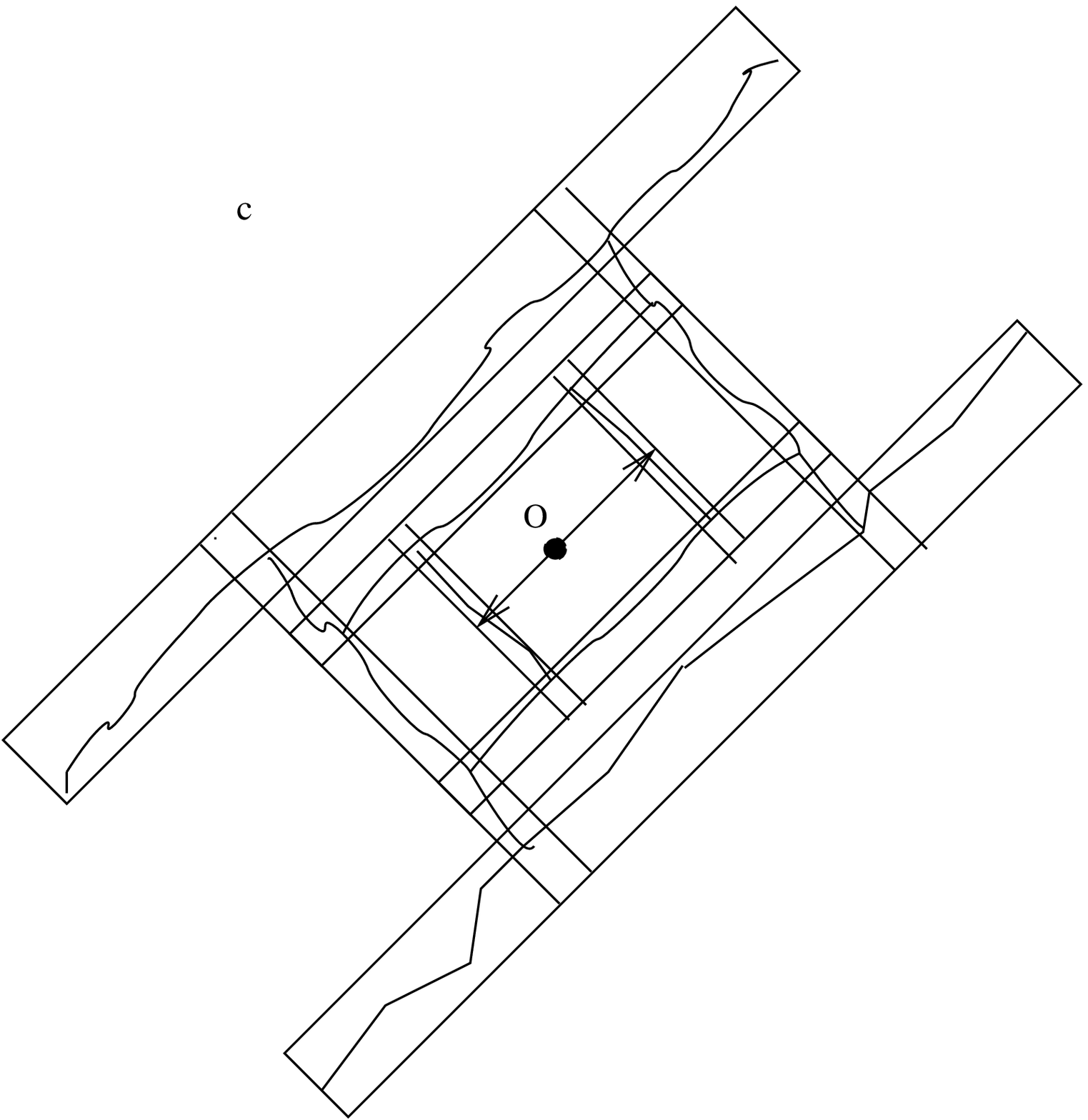}
}
\caption{a) The sequence of intersecting rectangles. b) Dotted non straight line stand for NE-SW (NW-SE) clusters spanning the rectangles
 c) Frozen structure containing the origin}
\label{disc}
\end{figure}

\subsection{Crossover length}

%In the previous sections we have analyzed the behavior of SM
%in the thermodynamic limit. 
Let us now focus on the divergence of the incipient blocked cluster
approaching the transition. This can be studied analyzing 
the typical size,  $L_c$, below which frozen clusters
occur on finite lattices. 

% diverges faster than any power law
%when $p\nearrow p_c^{DP}$: $\log
%L_c(p)\simeq k (p-p_c^{DP})^{-\mu}$, where
%$\mu=\nu_{\parallel}(1-1/z)\simeq 0.64$.

%Also, as for BP, %it is again possible to verify that 
%$L_c$ 
%has a meaning on the infinite lattice: $L_c^2$ gives the typical
%time needed to empty the origin.
%dire in che senso incipient
%In the following  we will show
%that $L_c$    

%We sketch separately the
%arguments leading to lower and upper bounds for $L_c$.

%leading to this result.
%$\log L_c(p)\geq k
%(p-p_c^{DP})^{-\mu}$ and $\log L_c(p)\leq k (p-p_c^{DP})^{-\mu}$ with
%$k_l,k_u$ two positive constants.

%To establish the lower bound we use again the
% T-junctions and
%construct a set of configurations
% containing a frozen backbone whose probability
% goes to one when $p\nearrow p_c^{DP}$ and $L\to\infty$ with
% $\log L\leq k_l(\rho-\rho_{dp})^{-\mu}$. 
We first obtain a lower bound on $L_c$ constructing explicitly 
blocked structures that exist with finite probability as long as 
$L<L_c^{lb}$.  Consider 
 NE-SW and NW-SE paths of length $s$ intersecting as in
 Fig.3 b).
%\ref{Qell} b)
This type of structure can be
 emptied completely 
only starting from its border since each finite directed path
 terminates on both ends into T-junctions with a path in the transverse
 direction. Therefore it is frozen
 if we continue the construction up to the
 border of the lattice and we  take periodic 
 boundary conditions. 
%and we consider periodic boundary conditions, 
% Furthermore, a similar frozen backbone exists
% also if one (or more) of the finite paths is displaced inside an adjacent
% rectangular region of size $s\times s/6$, as shown in Fig.
% \ref{Qell}. 
Thus the probability that there exists a frozen
 cluster, $1-R(L,p)$, is bounded from below by the
 probability that {\sl each} of the $O(L/s)^2$ dotted rectangles in Fig. 3 b)
 contains at least one
 path connecting its short sides.  This leads to
$R(L,p)\leq (L/s)^2\exp(-cs^{1-z})$
provided $s\leq \xi_{\parallel}$ (for $s>\xi_{\parallel}$ the probability 
of having a DP cluster in a rectangle starts to decrease and cannot 
be bounded anymore by $1-O[\exp(-c\ell_i^{1-z})]$). Thus
%since the probability of not having
%a DP path in a region $s\times s/6$ is  $\simeq \exp(-cs^{1-z})$ as long as $s\leq\xi_{\parallel}$ 
%and, as a consequence, 
taking 
%the largest possible $s$ for which the previous 
%statements hold\footnote{For $s>\xi_{\parallel}$ the probability 
%of having a DP cluster in a rectangle starts to decrease and cannot 
%be bounded anymore by$1-O[\exp(-c\ell_i^{1-z})]$.
%} (
$s\propto \xi_{\parallel}$, we get
%$\lim_{ L\to\infty, p\nearrow p_c^{DP}}R(L,\rho)=0$ for 
%$L/\xi_{\parallel}\exp(c\xi_{\parallel}^{1-z})\rightarrow 0$, therefore
$\log L_c \geq k_l|p-p_c^{DP}|^{-\mu}$.

In order to establish an upper bound on $L_c$, we
determine the size $L^{up}_c$ above which
unstable voids, that can be expanded until emptying the
whole lattice, occur typically. 
The results in Section \ref{Absence} imply that
the probability of expanding an empty nucleus to infinity is
dominated by the probability  of expanding it up to
$\ell=\xi_{\parallel}$. Indeed, above this size the probability of an
event which prevents expansion is exponentially suppressed. Therefore,
considering the $O(L/\xi_{\parallel})^2$ possible positions for the region
that it is guaranteed to be emptyable up to size $\xi_{\parallel}$, we
can bound the probability that a lattice of linear size $L$ is emptyable 
 as $R(L,p)\geq L^2\delta$, where $\delta$
is the probability 
that a small empty nucleus can be expanded until size 
$\xi_{\parallel}$.
In the emptying procedure described in Section \ref{Absence} we evaluated the
cost for expanding of one step the empty region $Q_{\ell}$. 
Analogously, the cost
of expanding directly from $Q_{\ell}$ to $Q_{2\ell}$ can be
 bounded from below by  
$C^{{\ell}^{1-z}}$, with $C$ a positive constant independent from $\ell$.
This can be done by dividing the region contained in $Q_{2\ell}$ and not in $Q_{\ell}$ into $\ell^{1-z}$ strips with parallel and transverse length of order $\ell$ and $\ell^z$, requiring that none of them contains a DP path which percolates in the transverse direction and using for this event
the scaling hypothesis of directed percolation when $p\nearrow p_c^{DP}$.
Thus for the expansion up to size $\xi_{\parallel}$ we get
 $\delta\geq\prod_{i=1}^{\log_2 \xi_{\parallel}}C^{2^{i(1-z)}}\simeq
\exp(-C'\xi_{\parallel}^{1-z})$, with $C'>0$ .
 This, together with above inequality, yields 
$\log L_c \leq k_u|p-p_c^{DP}|^{-\mu}$.

As a consequence upper and lower bound leads to the same scaling 
at leading order and imply that the cross-over length diverges 
with an essential singularity, i.e. faster
than any power law for $p\nearrow p_c^{DP}$.

\section{Discussion}
Let us first discuss the dynamical behavior of the SM model. 
The results of the previous sections have important consequence on the
dynamics of SM. First, incipient blocked clusters can be unblocked only from
the boundary. As a consequence the relaxation timescale is expected to 
scale at least as (but likely larger than) their 
typical size: $\propto \exp (k/|p-p_c^{DP}|^{\mu})$. 
Indeed this can be proved rigorously \cite{TRoma}.
Furthermore, since the 
fraction of blocked sites is finite at the transition, two point dynamical 
correlation functions, e.g. spin-spin correlations, will show a plateau
like super-cooled liquids approaching the glass transition. The plateau, 
also called Edwards-Anderson parameter in the context of spin-glasses, 
corresponds to the frozen spin fluctuations. 
These two dynamical characteristics are remarkable since they lead to a 
dynamics qualitatively similar to the one of glass-forming liquids. 
It would be very interesting to perform more detailed investigations 
and comparisons, in particular by numerical simulations.

%***********MI SONO FERMATO QUI. BISOGNA AGGIUNEGERE SULLA UNIVERSALITA
%+ACKNOWLEDGEMENTS+occuparsi delle figs******

The extension and universality of the jamming percolation transition
of SM remain  fundamental questions to be investigated. As
it has been discussed in \cite{TBFcomment} (see also \cite{JScomment,JS}), 
it is possible to identify
a class of rules which give rise to a jamming transition 
and belong to the same universality class of SM: as $p\nearrow p_c$
 the divergence of the incipient frozen cluster
follows the same scaling and the transition remain discontinuous. 
For all these models the
jamming transition is a consequence of the existence 
of (at least) two transverse
 directed percolation (DP)-like processes which can form a 
network of finite DP-clusters blocked by T-junctions with clusters in the transverse direction. 
A model that belongs to this class 
is for example 
the Knight model defined in \cite{letterTBF}. 
Note that in general, at variance with SM, it will not be possible
to determine analytically the exact value of $p_c$ (this was possible for SM thanks to the fact that in each of the two transverse directions the blocked clusters can be exactly mapped into those of conventional 2 dimensional DP).
Neither it is possible to generalize the rigorous proofs obtained for the SM. 
However, it is nevertheless possible to obtain numerically 
a reliable estimate of $p_c$ and a confirmation that the transition has the same properties of SM. 
This is done analyzing finite size effects with proper choices of the
%for the corresponding cellular automaton 
geometry and boundary conditions which allow to focus separately on each 
of the two transverse directions. In this way one can verify that on 
long length scales the two independent directional processes are in the DP universality 
class and obtain a good numerical estimate of $p_c$. Using suitable boundary conditions and geometries 
% and give a proper receipt
%which allows to obtain a reliable estimate of $p_c$ from numerical
is particularly important for jamming percolation since, as for bootstrap percolation,
convergence to the asymptotic results can be extremely slow.
For an extended discussion on this
we refer 
to \cite{TBFcomment}, where the value
of the critical density for the Knight model has been derived. 
%This was done by performing simulations with proper choices of the
%for the corresponding cellular automaton 
%geometry and the boundary conditions which allow to focus separately into each of the two transverse directions, i.e.
% to avoid
%blocked clusters formed by finite DP-like processes ended by T-junctions
%in the transverse direction and thus 
%to study separately 
%the two DP-like processes for which the usual finite size scaling tools can be applied to derive $p_c$ (since their transition is second order). 
The result is
$p_c^{Knight}\simeq 0.635$ and differs 
from our original conjecture $p_c^{Knight}=p_c^{DP}$ \cite{letterTBF} which
was due to the overlooking of some blocked structures 
 \cite{JScomment}.

We thank D.S.Fisher for very important collaborations on this subject.

% BibTeX users please use
% \bibliographystyle{}
% \bibliography{}
%
% Non-BibTeX users please use

\end{document}